\documentclass[
 superscriptaddress,
 reprint,
 amsmath, amssymb,
 aps, prd, preprintnumbers, nofootinbib, floatfix
]{revtex4-2}

\usepackage{graphicx}
\usepackage{dcolumn}
\usepackage{bm}
\usepackage{color}
\usepackage{amssymb}
\usepackage{amsmath}
\usepackage{braket}
\usepackage[colorlinks=true,allcolors=Blue,pdfusetitle]{hyperref}
\usepackage{rotating}
\usepackage{multirow}
\usepackage{graphicx}
\usepackage[dvipsnames]{xcolor}

\usepackage{tikz}
\usetikzlibrary{quantikz}

\usepackage{svg}
\usepackage{esdiff}
\definecolor{Green}{RGB}{0, 128, 0}
\newcommand{\orcid}[1]{\href{https://orcid.org/#1}{\includegraphics[width=10pt]{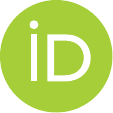}}}

\begin{document}
\title{Collective neutrino oscillations on a quantum computer with hybrid quantum-classical algorithm}

\author{Pooja Siwach \orcid{0000-0001-6186-0555}}
\email{siwach1@llnl.gov}
\affiliation{Nuclear and Chemical Science Division, Lawrence Livermore National Laboratory, Livermore, California 94551, USA}
\affiliation{
Department of Physics, University of Wisconsin--Madison,
Madison, Wisconsin 53706, USA}

\author{Kaytlin Harrison}
\email{kfharrison@wisc.edu}
\affiliation{
Department of Physics, University of Wisconsin--Madison,
Madison, Wisconsin 53706, USA}

\author{A. Baha Balantekin \orcid{0000-0002-2999-0111}}
\email{baha@physics.wisc.edu}
\affiliation{
Department of Physics, University of Wisconsin--Madison,
Madison, Wisconsin 53706, USA}

\date{\today}


\begin{abstract}
We simulate the time evolution of collective neutrino oscillations in two-flavor settings on a quantum computer. We explore the generalization of Trotter-Suzuki approximation to time-dependent Hamiltonian dynamics. The trotterization steps are further optimized using the Cartan decomposition of two-qubit unitary gates $U\in SU(4)$ in the minimum number of controlled-NOT (CNOT) gates making the algorithm more resilient to the hardware noise. A more efficient hybrid quantum-classical algorithm is also explored to solve the problem on noisy intermediate-scale quantum (NISQ) devices.

\end{abstract}

\maketitle

\section{Introduction}
\label{sec:introduction}

Neutrinos are dominant carriers of energy released during the core-collapse supernova explosion. Understanding the dynamics of this explosion would be greatly helped by a complete treatment of neutrino transport. The very large number of neutrinos emitted within tens of seconds during supernovae explosion stream out of the star interacting with each other and the surroundings. This problem cannot be modelled exactly within the full many-body picture because of the computational limitations. However, certain approximations can be made to simplify the problem [see e.g. Ref.~\cite{Balantekin:2006tg}]. These approximations may capture much of the physics of the neutrino transport, but to ensure that we are not missing any subtle issues, it would be beneficial to evaluate the many-body evolution of the system for a relatively large number of neutrinos. However, as the number of neutrinos increase, conventional methods of solving the time-dependent problems like the fourth order Runge-Kutta (RK4) method, rapidly become obsolete~\cite{Pehlivan:2011hp, Pehlivan:2014zua, Birol:2018qhx, Cervia:2019nzy, cervia:2019, Patwardhan:2019zta, Rrapaj:2019pxz, Patwardhan:2021rej, Lacroix:2022krq}. The tensor network methods can accommodate larger many-body systems but only up to a few tens of neutrinos which is also highly dependent on the initial state~\cite{Cervia:2022pro}. These simulations become more complicated when a more realistic scenario with three active flavors is considered~\cite{siwach:2023prd}. 

It is generally assumed that the quantum simulations are a more natural way to simulate the quantum many-body system. The simulations for the collective neutrino oscillations on quantum computers have already been attempted ~\cite{Hall:2021rbv, Yeter-Aydeniz:2021olz, Jha:2021itm, Illa:2022zgu, Amitrano:2022yyn}. Due to the hardware limitations, only a few-neutrino systems could be simulated in these studies. These studies consider a two-beam model for the collective neutrino oscillations.  In the present work, we are interested in a model where each neutrino occupies a different energy bin. We then have to simulate a time-dependent many-body problem on the quantum computer. 

Several quantum algorithms, from pure quantum to variational principle based hybrid quantum-classical algorithms, have been devised for simulating the time-dependent Hamiltonian dynamics on a quantum computer~\cite{Yuan:2019quantum,Benedetti:2021prr,Heya:2019arx,Cirstoiu:2020npj,Commeau:2020arx,Gibbs:2021arx,lau2021noisy,Haug:QST2022,Kishor:RMP2022,otten:2019arx,Barison:2021quantum}. The most commonly used quantum algorithms for the dynamics of time-dependent Hamiltonian are summarized in Ref.~\cite{lau2021noisy} with the details on circuit complexity and Barren plateau. In summary, the pure quantum algorithm, like Trotterization, require very deep quantum circuits with a large number of quantum gates. On the other hand, the variational methods like variational quantum simulations (VQS)~\cite{Li:2017prx,Yuan:2019quantum,Endo:prl2020} and variational fast forwarding (VFF)~\cite{Cirstoiu:2020npj} may face the Barren plateau problem. The quantum assisted algorithm, which we are using in the present work, is a hybrid quantum-classical algorithm and it requires the quantum computer only once, that means there is no classical feedback loop. Therefore, the Barren plateau problem does not occur and the circuit complexity is significantly reduced.

We are interested in simulating the dynamics of collective neutrino oscillations under a time-dependent Hamiltonian on a quantum computer. For this purpose, we first explore the generalization of Trotter-Suzuki method, which does not have any classical feedback loop, for the time-dependent Hamiltonian. To reduce the circuit complexity, we incorporate the Cartan decomposition of two-qubit gates in terms of minimum number of CNOT gates. For the utility of current noisy quantum computers, we utilise the hybrid quantum-classical algorithm based on the quantum assisted algorithm (QSA).


This paper is organised as follows. We begin by giving a  brief description of the neutrino Hamiltonian which describes the collective neutrino oscillations in section~\ref{sec:Hamiltonian}. The quantum algorithms employed to simulate the time-evolution of the Hamiltonian are given in section~\ref{sec:algortihm}. In section~\ref{sec:results}, we discuss the results obtained with both of the quantum algorithms and compare them with the classical results. We conclude the paper  briefly sketching out possible potential extensions in section~\ref{sec:conclusions}.





 
\section{The neutrino Hamiltonian}
\label{sec:Hamiltonian}

In general, it is considered that the flavor evolution of neutrinos depends on the vacuum oscillations, the interaction of neutrinos with the background matter (Mikheyev-Smirnov-Wolfenstein (MSW) matter effects), and the neutrino-neutrino ($\nu$-$\nu$) interaction. In the present work, similar to Refs.~\cite{Patwardhan:2019zta,cervia:2019,Cervia:2022pro}, we consider the settings where the $\nu$-$\nu$ interaction is dominant such that the MSW matter effects can be ignored. Therefore, in the mass basis, the total Hamiltonian of the system considered can be written as~\cite{cervia:2019}
\begin{equation}
    H= H_{\nu} + H_{\nu\nu}=-\sum_{\omega} \omega J_{\omega}^{z}+\mu(t)\sum_{\omega,\omega',\omega\ne\omega'}\Vec{J}_{\omega}\cdot\Vec{J}_{\omega'},
\end{equation}
where first and second terms correspond to the vacuum oscillations and the $\nu$-$\nu$ interaction, respectively. Here, $\omega$ represents the vacuum oscillation frequency and $\mu(t)$ accounts for the $\nu$-$\nu$ interaction strength which is time-dependent and averaged over angles between the momenta of interacting neutrinos~\cite{Cervia:2022pro}. The isospin operators $\Vec{J}_{\omega}$ are half of the Pauli spin matrices $\sigma_{\omega}=X,Y,Z$; $\Vec{J}_{\omega}=\sigma_{\omega}/2$.

We consider the discrete, equally spaced vacuum frequencies such that $\omega_{i}=i\omega_{0}$, where $\omega_{0}$ is an arbitrary reference frequency. Therefore, replacing the isospin operators with the Pauli spin matrices and transforming to the flavor basis, the Hamiltonian for an $N$ neutrino system becomes
\begin{eqnarray}\label{eq:Hamiltonian}
    H &=& \frac{1}{2}\left[\sum_{i=0}^{N-1}(i+1)(\sin\theta X_{i}-\cos\theta Z_{i})\right.\nonumber\\
    &~&\left.+\mu(t)\sum_{i=0}^{N-1}\sum_{j>i}^{N-1}\left(X_{i}X_{j}+Y_{i}Y_{j}+Z_{i}Z_{j}\right)\right].
\end{eqnarray}
To separate out the Hamiltonian's time-independent and time-dependent parts, we write the Hamiltonian as
\begin{equation}\label{eq:Hamiltonian_split}
    H(t)=H_{I}+\mu(t)H_{D}
\end{equation}
where $H_{I}$ and $H_{D}$ represent the vacuum oscillations (time-independent) and the $\nu$-$\nu$ interaction (time-dependent), respectively. The strength of the interaction, $\mu$, provides the time-dependence. 
\section{Quantum algorithm for time-evolution}\label{sec:algortihm}
In order to understand the dynamics of the system under the Hamiltonian given in Eq.~\eqref{eq:Hamiltonian}, the following time-dependent Schr\"odinger equation has to be solved.
\begin{equation}\label{eq:time-dependent Sch eq}
    i\frac{d}{dt}\ket{\psi(t)}=H(t)\ket{\psi(t)}
\end{equation}
where $H(t)$ is the time-dependent Hamiltonian given in Eq.~\eqref{eq:Hamiltonian}. The solution to the above Eq.~\eqref{eq:time-dependent Sch eq} is given by
\begin{eqnarray}\label{eq:solution_of_TDSE}
    \ket{\psi(t+\Delta t)}=\mathcal{T}\left[\exp\left(-i\int_{t}^{t+\Delta t}H(t')dt'\right)\right]\ket{\psi(t)}
\end{eqnarray}
where $\mathcal{T}$ is the time-ordering operator. For a  time-independent Hamiltonian $H(t)=H,~~\forall t$, Eq.~\eqref{eq:solution_of_TDSE} becomes
\begin{equation}\label{eq:solution_of_TISE}
    \ket{\psi(t+\Delta t)}=\exp\left(-iH\Delta t\right)\ket{\psi(t)}.
\end{equation}
To perform the quantum simulations of time-evolution of many neutrino system under the time-dependent Hamiltonian $H(t)$ given in Eq.~\eqref{eq:Hamiltonian}, we explore three algorithms in this work.

\subsection{Trotterization formula for a time-dependent Hamiltonian}\label{subsec:trotter}

For simplicity, let us consider a time-independent Hamiltonian with the solution of corresponding Schr\"odinger equation given in Eq.~\eqref{eq:solution_of_TISE}. The exponential in this equation is difficult to compute especially when the Hamiltonian has several non-commutative terms. Therefore, an approximate solution in terms of product of exponentials is required for solving it on quantum computers. The Trotterization~\cite{Trotter:1959pams,Suzuki:1991jmp} is a useful procedure in that direction; for example, if $H= \sum_{k}H_{k}$, where $H_{k}$ are non-commutative operators, then the exponential given in Eq.~\eqref{eq:solution_of_TISE} under first-order Trotterization becomes
\begin{equation}\label{eq:trotter}
    e^{-i\Delta t\sum_{k}H_{k}}=\prod_{k}e^{-i\Delta t H_{k}}+\mathcal{O}[(\Delta t)^{2}].
\end{equation}
To compute the exponential of Hamiltonian approximately, we can implement Eq.~\eqref{eq:trotter} such that
\begin{equation}\label{eq:trotter_app}
    e^{-i\Delta t\sum_{k}H_{k}}\approx\prod_{k}e^{-i\Delta t H_{k}} \,
\end{equation}
and the solution is accurate up to error of order $\mathcal{O}[(\Delta t)^{2}]$. Higher-order formulae for Trotterization reviewed in Ref.~\cite{Hatano2005} can further reduce the order of errors.

The above mentioned Trotterization approach can be generalized to the time-dependent Hamiltonian case. However, only a few algorithms have been devised for this case~\cite{Huyghebaert_1990,Hatano2005,ikeda2023minimum}.
If $H(t)= \sum_{k}H_{k}(t)$, the time evolution operator given in Eq.~\eqref{eq:solution_of_TDSE} approximately becomes~\cite{Hatano2005}
\begin{equation}\label{eq:Trotter_TD}
    \mathcal{T}\left[\exp\left(-i\int_{t}^{t+\Delta t}H(t')dt'\right)\right]\approx\prod_{k}\exp(-i \Delta t H_{k}(t+\Delta t)).
\end{equation}

Utilizing the product formulae given in Eqs.~\eqref{eq:trotter_app} and \eqref{eq:Trotter_TD}, we can write the Trotterization formula for Hamiltonian corresponding to collective neutrino oscillations given in Eq.~\eqref{eq:Hamiltonian_split}.
\begin{eqnarray}\label{eq:exp_final}
    &\mathcal{T}&\left[\exp\left(-i\int_{t}^{t+\Delta t}\left(\sum_{k}H_{I}^{k}+\mu(t')\sum_{m}H_{D}^{m}\right)dt'\right)\right]\nonumber\\
    &\approx&\prod_{k}\exp{(-i\Delta t H_{I}^{k})}\prod_{m}\exp(-i \Delta t \mu(t+\Delta t)H_{D}^{m}).
\end{eqnarray}

\subsubsection{Quantum circuit with brute force}
To simulate the neutrino system, we have to design the quantum circuits corresponding to Eq.~\eqref{eq:exp_final} to be implemented on the quantum computer. For illustration, let us consider a two-neutrino system. The corresponding Hamiltonian derived from Eq.~\eqref{eq:Hamiltonian} can be written as
\begin{eqnarray}\label{eq:ham_two_neutrino}
    H &=& \frac{1}{2}\left[\sin\theta X_{0}-\cos\theta Z_{0}+2\sin\theta X_{1}-2\cos\theta Z_{1}\right.\nonumber\\
    &~&+\left.\mu(t)(X_{0}X_{1}+Y_{0}Y_{1}+Z_{0}Z_{1})\right]
\end{eqnarray}
Utilizing Eq.~\eqref{eq:Trotter_TD}, and using the short-hand notations for $\sin\theta=s$ and $\cos\theta=c$, the time-evolution operator can be written as
\begin{eqnarray}\label{eq:unitary_two_neutrino}
    U(t+\Delta t, t)&=&e^{-\frac{i}{2}s\Delta t X_{0}}e^{\frac{i}{2}c\Delta t Z_{0}}e^{-is\Delta t X_{1}}e^{ic\Delta t Z_{1}}\nonumber\\
    &~&\times e^{-\frac{i}{2}\mu(t+\Delta t)\Delta t X_{0}X_{1}}e^{-\frac{i}{2}\mu(t+\Delta t)\Delta t Y_{0}Y_{1}}\nonumber\\
    &~&\times e^{-\frac{i}{2}\mu(t+\Delta t)\Delta t Z_{0}Z_{1}}.
\end{eqnarray}

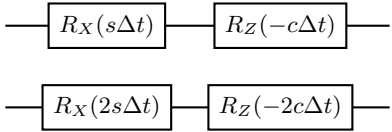
\begin{figure}[h!]
\begin{tikzpicture}
\node[scale=0.99] {
\begin{quantikz}
\qw&\gate{R_{X}(s\Delta t)}&\gate{R_{Z}(-c\Delta t)}&\qw{}\\
\qw&\gate{R_{X}(2s\Delta t)} & \gate{R_{Z}(-2c\Delta t)}& \qw
\end{quantikz}
};
\end{tikzpicture}
\caption{The quantum circuit corresponding to the one-body time-independent term in the two-neutrino Hamiltonian given in Eq.~\eqref{eq:unitary_two_neutrino}.}\label{fig:one-body}
\end{figure}

\begin{figure}[h!]
\begin{tikzpicture}
\node[scale=0.9, label=below:(a)] {
\begin{quantikz}
\qw&\ctrl{1}&\qw&\ctrl{1}&\qw{}\\
\qw&\targ{} & \gate{R_{Z}(\mu'\Delta t)}& \targ{}& \qw
\end{quantikz}
};
\end{tikzpicture}
\begin{tikzpicture}
\node[scale=0.9, label=below:(b)] {
\begin{quantikz}
\qw&\gate{H}&\ctrl{1}&\qw&\ctrl{1}&\gate{H}&\qw{}\\
\qw&\gate{H}&\targ{} & \gate{R_{Z}(\mu'\Delta t)}& \targ{}&\gate{H}&\qw
\end{quantikz}
};
\end{tikzpicture}

\begin{tikzpicture}
\node[scale=0.9, label=below:(c)] {
\begin{quantikz}
\qw&\gate{S^{\dagger}}&\gate{H}&\ctrl{1}&\qw&\ctrl{1}&\gate{H}&\gate{S}&\qw{}\\
\qw&\gate{S^{\dagger}}&\gate{H}&\targ{} & \gate{R_{Z}(\mu'\Delta t)}& \targ{}&\gate{H}&\gate{S}&\qw
\end{quantikz}
};
\end{tikzpicture}
\caption{The quantum circuits corresponding to (a) $e^{-\frac{i}{2}\mu'\Delta t Z_{0}Z_{1}}$, (b) $e^{-\frac{i}{2}\mu'\Delta t X_{0}X_{1}}$ and (c) $e^{-\frac{i}{2}\mu'\Delta t Y_{0}Y_{1}}$ exponentials of two-body time-dependent term in the two-neutrino Hamiltonian given in Eq.~\eqref{eq:unitary_two_neutrino}, where $\mu'=\mu(t+\Delta t)$.}\label{fig:two_body}
\end{figure}
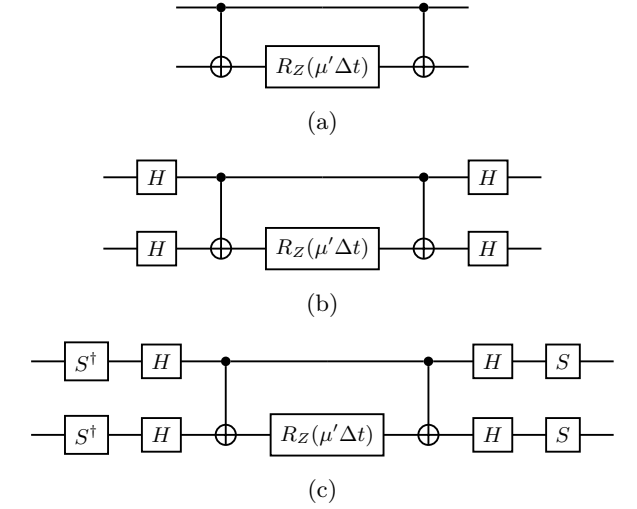

\subsubsection{Decomposition of two-qubit gates in minimal number of CNOTs}
\begin{figure}
\begin{tikzpicture}
\node[scale=0.68] {
\begin{quantikz}
\qw&\ctrl{1}&\gate{R_{X}(\mu'\Delta t)}&\gate{H}&\ctrl{1}&\gate{S}&\gate{H}&\ctrl{1}&\gate{R_{X}(\pi/2)}&\qw{}\\
\qw&\targ{}&\gate{R_{Z}(\mu'\Delta t)}&\qw&\targ{} & \gate{R_{Z}(-\mu'\Delta t)}&\qw& \targ{}&\gate{R_{X}(-\pi/2)}&\qw
\end{quantikz}
};
\end{tikzpicture}
\caption{The quantum circuit corresponding to $e^{-\frac{i}{2}\mu'\Delta t (X_{0}X_{1}+Y_{0}Y_{1}+Z_{0}Z_{1}})$  exponential of two-body time-dependent term in the two-neutrino Hamiltonian given in Eq.~\eqref{eq:unitary_two_neutrino}, in terms of minimum number of CNOT gates.}\label{fig:Cartan}
\end{figure}
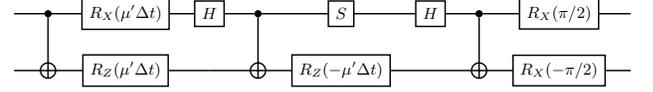

As can be seen in the figures~\ref{fig:one-body} and \ref{fig:two_body}, the total number of $19$ single-qubit gates and $6$ CNOT gates are required to simulate the two-neutrino Hamiltonian for a single time-step. The one-body term requires only $4$ single-qubit gates and the remaining $15$ single-qubit and $6$ CNOT gates are required for the two-body interaction term $X_{0}X_{1}+Y_{0}Y_{1}+Z_{0}Z_{1}$. The two-body term can be simulated by decomposing the corresponding two-qubit unitary gate $e^{-\frac{i}{2}\mu(t+\Delta t)\Delta t (X_{0}X_{1}+Y_{0}Y_{1}+Z_{0}Z_{1})}$ in terms of minimal CNOT gates. As derived in Refs.~\cite{Vidal:2004pra,Coffey:2008pra}, an arbitrary two-qubit unitary gate $U\in SU(4)$ can be decomposed into a quantum circuit with a maximum of $3$ CNOT gates and four layers of single-qubit gates. The quantum circuit corresponding to the two-body interaction term of neutrino Hamiltonian under this scheme is given in Figure~\ref{fig:Cartan}. One can see that for a two neutrino case, it requires $3$ CNOT gates and $8$ single-qubit gates for two-body interaction term, reducing them by a factor of nearly half. Therefore, under this scheme, the quantum simulation become more resilient to the gate errors and hardware noise.

\subsection{Quantum assisted simulator: hybrid quantum-classical algorithm}\label{sec:QAS}
As can be noticed from the trotterization algorithm given in previous subsection, a large number of single and two-qubit gates are required to perform quantum simulations even for a small system of two neutrinos. Decomposing the two-qubit gates in the minimum number of CNOTs reduces the circuit depth, but still the number of gates is very large to perform the simulations on the NISQ devices. Therefore, to harness the full potential of NISQ devices and perform the simulations with limited errors, we explore a hybrid quantum-classical algorithm based on the quantum assisted simulator (QAS)~\cite{Bharti:2021,lau2021noisy}.
The algorithm is well documented in Ref.~\cite{Bharti:2021}, but we present a minimal description here including the modification implemented in our work.

First, we explain the algorithm for time-independent Hamiltonian which can be easily extended to the time-dependent case. Assume that the Hamiltonian can be written as the linear combination of $r$ unitaries such that
\begin{equation}\label{eq:ham_unitary}
    H = \sum_{i=1}^{r}c_{i}U_{i}.
\end{equation}

In the QAS, we take the ansatz as time-dependent linear combination of $m$ basis states $\ket{\psi_{i}}$ given by
\begin{equation}\label{eq:phi_alpha}
    \ket{\phi(\alpha(t))}=\sum_{i=0}^{m-1}\alpha_{i}(t)\ket{\psi_{i}}\,
\end{equation}
where $\alpha_{i}(t)\in \mathbb{C}$. The normalization of the ansatz wave function, {\it i.e.,} $\braket{{\phi(\alpha)}|{\phi(\alpha)}}=1$, can be accomplished as
\begin{equation}
    \alpha^{\dagger}\mathcal{E}\alpha =1
\end{equation}
where $\mathcal{E}$ is the matrix having overlaps of basis states
\begin{equation}
    \mathcal{E}_{i,j}=\braket{\psi_{i}|\psi_{j}}.
\end{equation}
Using the Dirac and Frenkel variational principle, we get the final equation as
\begin{equation}\label{eq:ed}
    \mathcal{E}\frac{\partial\alpha(t)}{\partial t}=-i\mathcal{D}\alpha(t)\,
\end{equation}
where
\begin{equation}\label{eq:overlaps_Dij}
    \mathcal{D}_{i,j}=\sum_{k}\beta_{k}\bra{\psi_{i}}U_{k}\ket{\psi_{j}}.
\end{equation}
After solving for $\dot{\alpha}(t)$, we can update the parameters as
\begin{equation}
    \alpha_{j}(t+\delta t) = \alpha_{j}(t) + \dot{\alpha}_{j}(t)\delta t,
\end{equation}
and hence the wave function given in Eq.~\eqref{eq:phi_alpha}.

The most important step in QAS is choosing the ansatz. The selection of ansatz suggested in Ref.~\cite{Bharti:2021} is based on the cumulative $K$-moment states. The $K$-moment states are a set of quantum states which can be defined as $U_{i_{K}}\ldots U_{i_{2}}U_{i_{1}}\ket{\psi}$ where $U_{i_{l}}\in \mathbb{U}$ are the unitaries forming the Hamiltonian. The cumulative $K$-moment states are a union set of all, {\it i.e.,} $0$-moment, $1$-moment up to $K$-moment states. Therefore, in the 
case of time-independent Hamiltonian, the ansatz can be defined as the linear combination of cumulative $K$-moment states, which can be calculated by applying the unitaries $U_{i}$ of the Hamiltonian given in Eq.~\eqref{eq:ham_unitary}.

In the case of time-dependent Hamiltonian of the form
\begin{equation}\label{eq:time_dependent Hamiltonian}
    H(t)=H_{I}+f(t)H_{D},
\end{equation}
where $H_{I}=\sum_{k}\beta_{k}U_{k}$ and $H_{D}=\sum_{k}\gamma_{k}V_{k}$ are the linear combinations of unitaries $U_{k}$ and $V_{k}$, respectively, the only difference in the algorithm comes at the point of designing an ansatz. For the Hamiltonian of the form given in Eq.~\eqref{eq:time_dependent Hamiltonian}, the ansatz can be designed based on the effective form of Hamiltonian given in the theory of quantum control. The effective Hamiltonian can be written as linear combination of all nested commutators of $H_{I}$ and $H_{D}$. Such that
\begin{equation}\label{eq:effective_Hamiltonian}
    H_{\rm eff}=\sum_{i}c_{i}C_{i}
\end{equation}
where $C_{i}\in\{H_{I},[H_{I},H_{D}],[H_{I},[H_{I},H_{D}]],\ldots\}$. Similar to directly using $U_{i}$ in the case of time-independent Hamiltonian as explained above, we can use the $C_{i}$'s to form the ansatz for the time-dependent case. However, not all $C_{i}$s are unitary, and hence these can be further decomposed into the unitary terms forming a bigger set of basis states. Now, similar to Eq.~\eqref{eq:overlaps_Dij}, $\mathcal{D}_{ij}$ can be written as
\begin{equation}
    \mathcal{D}_{ij}=\sum_{k}\beta_{k}\bra{\psi_{i}}U_{k}\ket{\psi_{j}}+f(t)\sum_{k}\gamma_{k}\bra{\psi_{i}}V_{k}\ket{\psi_{j}}.
\end{equation}
\begin{figure}[h!]
\begin{tikzpicture}
\node[scale=0.9] {
\begin{quantikz}
\lstick{$\ket{0}$}&\gate{H}&\ctrl{1}&\gate{H}&\meter{}\\
\lstick{$\ket{\psi}$}&\qwbundle[alternate]{} & \gate{U}\qwbundle[alternate]{}& \qwbundle[alternate]{}& \qwbundle[alternate]{} 
\end{quantikz}
};
\end{tikzpicture}
\caption{The quantum circuit corresponding to Hadamard test for calculating Re$\bra{\psi}U\ket{\psi}$ which is the difference between measuring $0$ and $1$ on the ancilla qubit.}\label{fig:had}
\end{figure}
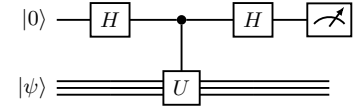
\begin{figure*}
    \centering
    \includegraphics[width=0.99\linewidth]{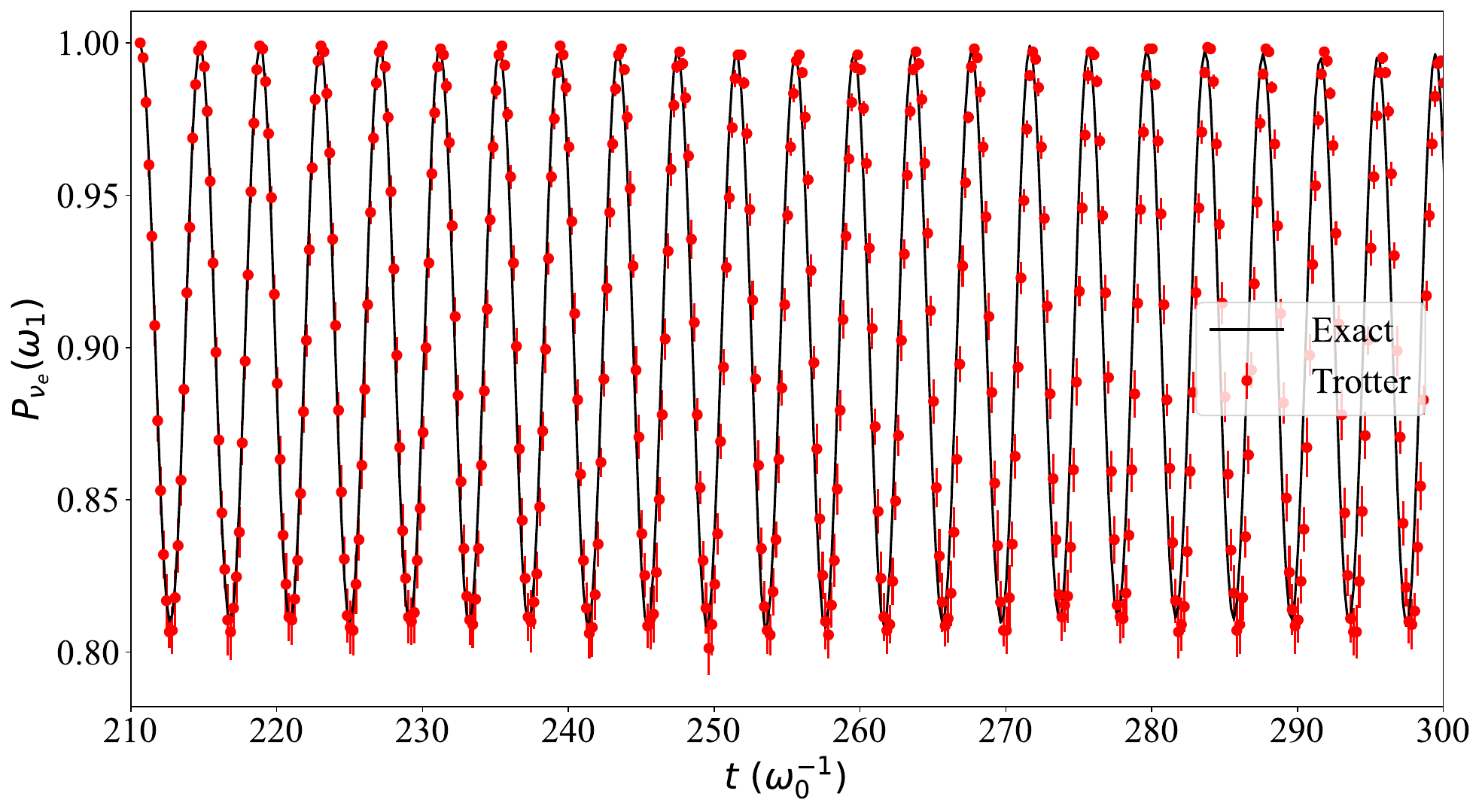}
    \caption{The survival probability for the neutrino in $\omega_{1}$ bin to be in the $\nu_{e}$ flavor in case of the two-neutrino system with the initial state $\ket{\nu_{e}\nu_{e}}$. Quantum simulation results with the trotterization for the time-dependent Hamiltonian are labeled as ``Trotter", and the classical results are labeled as ``Exact".}
    \label{fig:2N_trotter}
\end{figure*}

Therefore, we need to calculate the matrix elements $\mathcal{E}_{ij}$ and $\mathcal{D}_{ij}$ on the quantum computer only once. Then, Eq.~\eqref{eq:ed} can be solved on the classical computer. In this way, the simulations can be performed with minimal errors. The matrix element $\mathcal{E}_{ij}$ and $\mathcal{D}_{ij}$ can be conveniently calculated on the quantum computer using Hadamard test. The quantum circuit for the Hadamard test is shown in Figure~\ref{fig:had}. The $K$-moment matrix elements can be written as
\begin{equation}
    \mathcal{E}_{ij}=\braket{\psi_{i}|\psi_{j}}=\bra{\psi_{0}}U_{i_{1}}^{\dagger}\ldots U_{i_{K}}^{\dagger}U_{j_{K}}\ldots U_{j_{1}}\ket{\psi_{0}},
\end{equation}
and
\begin{equation}
    \mathcal{D}_{ij}=\bra{\psi_{i}}U_{n}\ket{\psi_{j}}=\bra{\psi_{0}}U_{i_{1}}^{\dagger}\ldots U_{i_{K}}^{\dagger}U_{n}U_{j_{K}}\ldots U_{j_{1}}\ket{\psi_{0}},
\end{equation}
such that the Hadamard test can be utilized for their computation on the quantum computer.

To illustrate the QAS algorithm for the time-dependent case, we consider the two-neutrino Hamiltonian given in Eq.~\eqref{eq:ham_two_neutrino}. For this Hamiltonian
\begin{equation}\label{eq:ham_two_neutrino_Hi}
    H_{I} = \frac{1}{2}\left[\sin\theta (X_{0}+2 X_{1})-\cos\theta (Z_{0}+2 Z_{1})\right],
\end{equation}
and 
\begin{equation}\label{eq:ham_two_neutrino_Hd}
    H_{D} = \frac{1}{2}\left[(X_{0}X_{1}+Y_{0}Y_{1}+Z_{0}Z_{1})\right].
\end{equation}

Following the above explained technique to calculate the set of permitted operators to design the ansatz, the obtained set contains the following operators
\begin{eqnarray}
    \mathcal{S}&=&\left\{I, X_{0},Z_{0},X_{1},Z_{1},X_{0}X_{1},Y_{0}Y_{1},Z_{0}Z_{1},\right.\nonumber\\
    &~&\left. Z_{0}Y_{1},Y_{0}Z_{1},Y_{0}X_{1},X_{0}Y_{1},Z_{0}X_{1},X_{0}Z_{1}\right\}
\end{eqnarray}
Even if we consider only 1-moment cumulative set to generate the basis states, using the entire set leads to $14$ basis states. Computing the $\mathcal{E}_{ij}$ and $\mathcal{D}_{ij}$ requires several operations to be performed on the quantum computer. Also, the number of coupled equations to be solved on the classical computers is large. We notice that the several operators lead to the same basis states except the phase. Therefore, we can use a smaller set of operators $\{I,X_{0},X_{1},X_{0}X_{1}\}$ only, to generate the basis. For an initial state $\ket{\psi_{0}}=\ket{00}$, we get
\begin{eqnarray}
\ket{\psi_{1}}&=&X_{0}\ket{\psi_{0}}\\
\ket{\psi_{2}}&=&X_{1}\ket{\psi_{0}}\\
\ket{\psi_{3}}&=&X_{0}X_{1}\ket{\psi_{0}}
\end{eqnarray}
We utilise these basis states to calculate the overlaps $\mathcal{E}_{ij}$ and $\mathcal{D}_{ij}$, and solve the corresponding set of differential equations. The results for survival probabilities are given in next section.

\begin{figure*}[h!]
    \centering
    \includegraphics[width=0.99\linewidth]{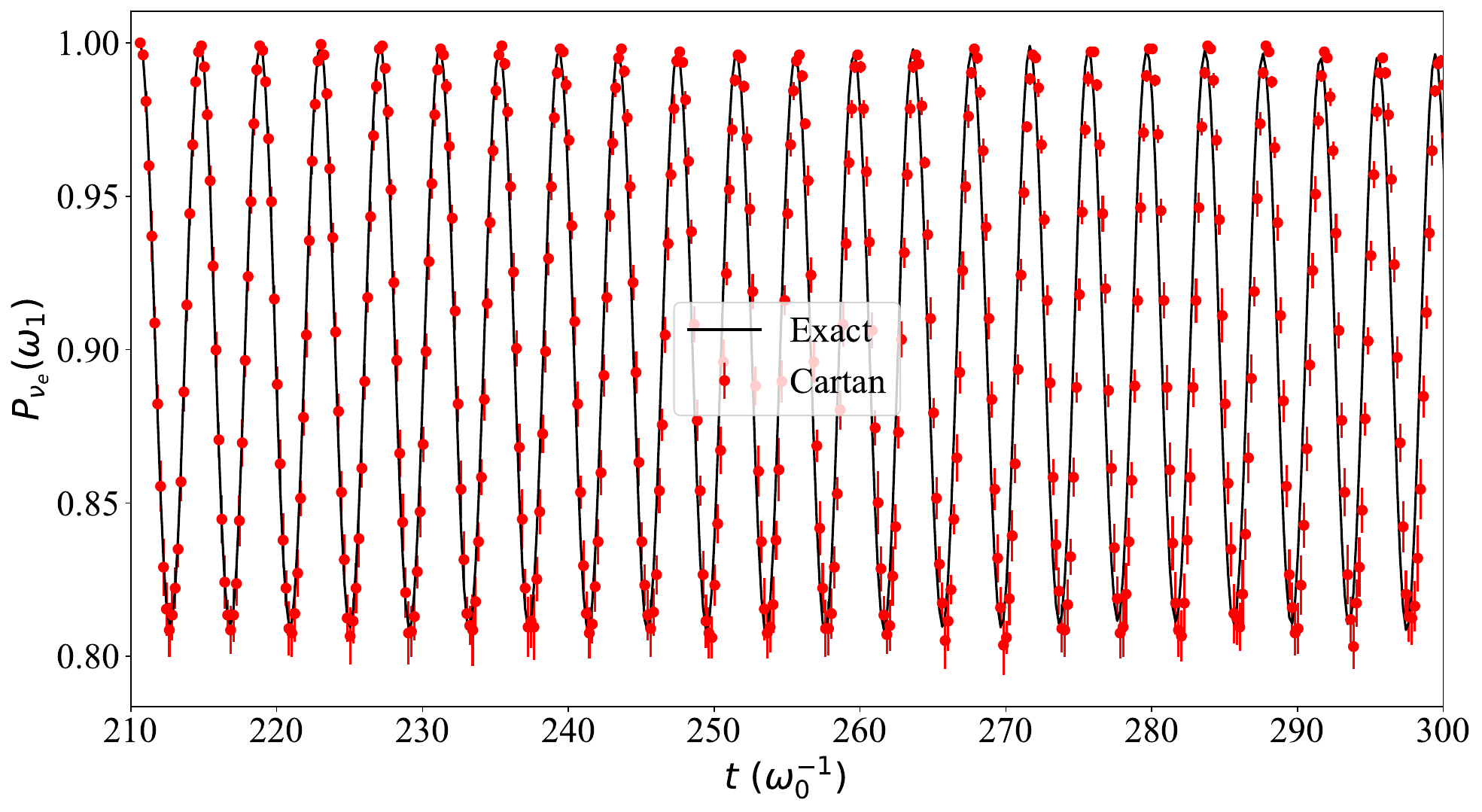}
    \caption{Same as Figure~\ref{fig:2N_trotter}, but with the quantum algorithm with the minimum number of CNOTs for two-qubit gates labeled as ``Cartan".}
    \label{fig:2N_cartan}
\end{figure*}
\begin{figure*}
    \centering
    \includegraphics[width=0.99\linewidth]{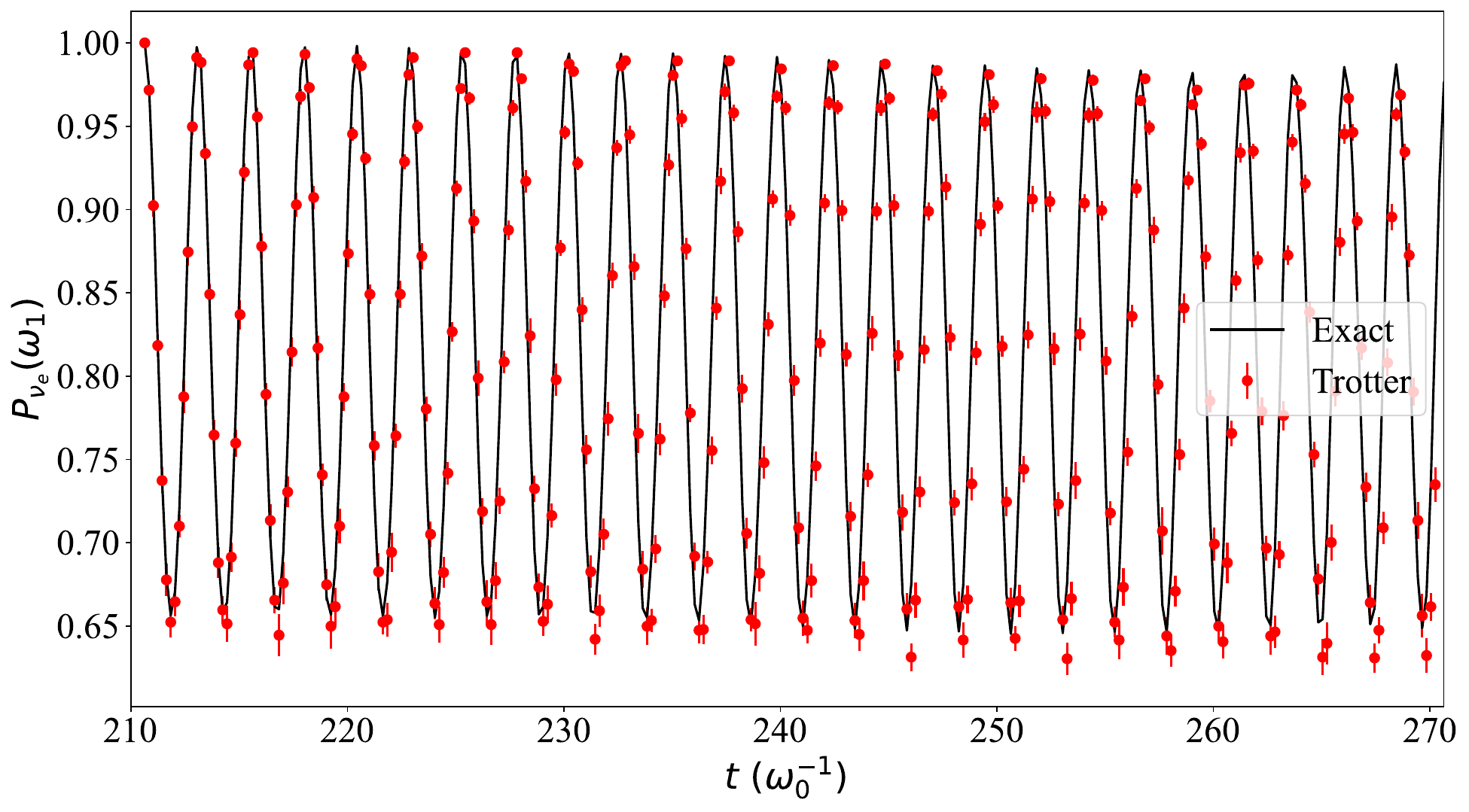}
    \caption{The survival probability for the neutrino in $\omega_{1}$ bin to be in the $\nu_{e}$ flavor in case of the four-neutrino system with the initial state $\ket{\nu_{e}}^{\otimes 4}$. Quantum simulation results with the trotterization for the time-dependent Hamiltonian are labeled as ``Trotter", and the classical results are labeled as ``Exact".}
    \label{fig:4N_trotter}
\end{figure*}
\begin{figure*}
    \centering
    \includegraphics[width=0.99\linewidth]{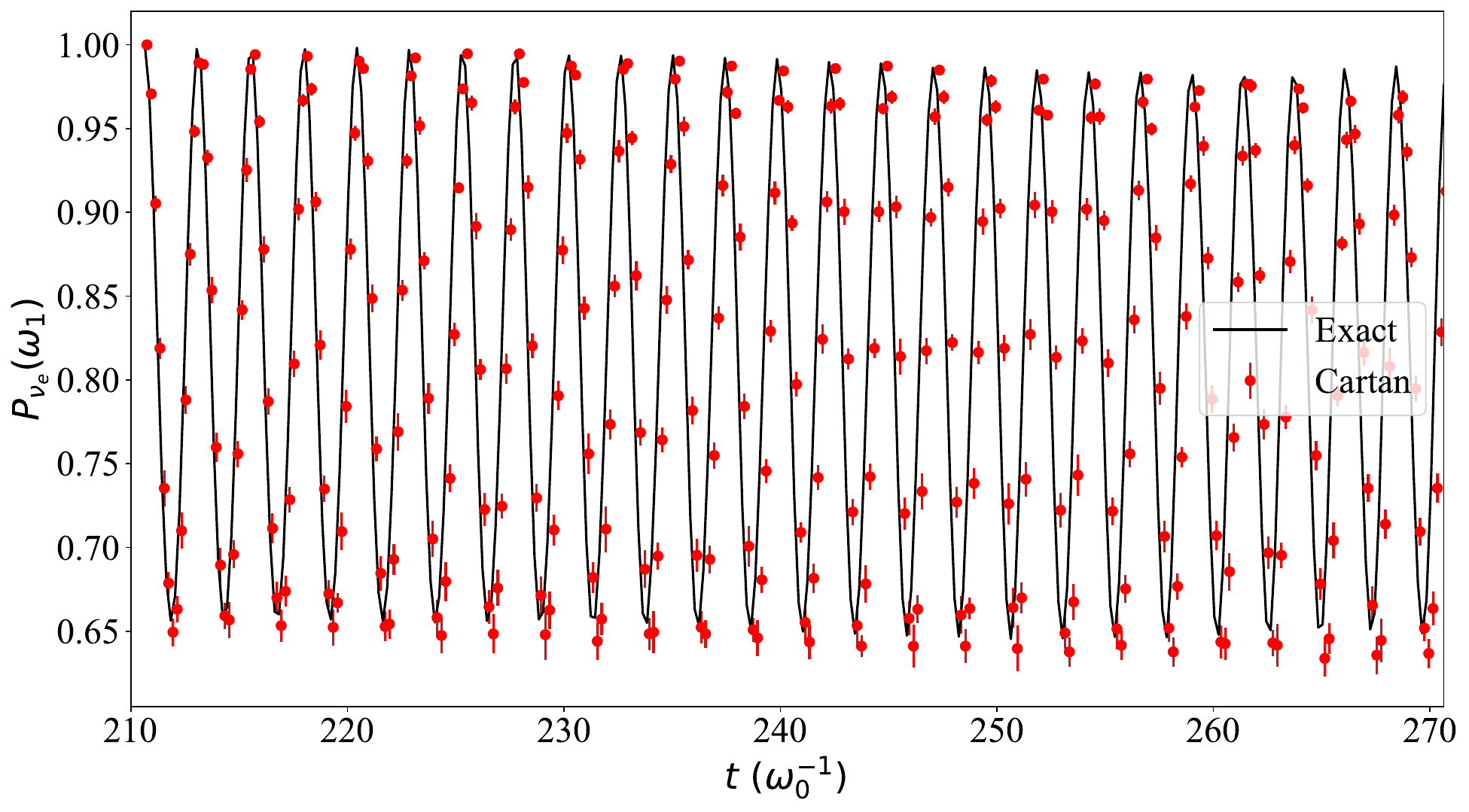}
    \caption{Same as Figure~\ref{fig:4N_trotter} but with the Cartan decomposition of two-qubit gates in terms of minimum number of CNOT gates. The quantum simulation results are labeled as ``Cartan" and the classical ones are labeled as ``Exact."}
    \label{fig:4N_cartan}
\end{figure*}
\section{Results}\label{sec:results}
We perform the quantum simulations for the two-neutrino and four-neutrino systems utilising all the algorithms described above.  We take the time-step as $0.2$, and the number of shots as $1024$ for all sets of calculations. For every time-step, $50$ runs of calculations are performed, and the median and median absolute deviation (MAD) are calculated for the estimation of exact value and the error bars, respectively. We do all the calculations on the IBM's QASM simulator. We start our simulations at $t=210.64\omega_{0}^{-1}$.

In the case of two-neutrino system, the results with the trotterization for the time-dependent Hamiltonian labeled as ``Trotter" are given in Figure~\ref{fig:2N_trotter}. At initial times, since the quantum circuits are shallow, the quantum simulation results match exactly with the classical results (``Exact") with negligibly small error bars. However, with the increase in time, the results start deviating more from the classical results. These deviations can be understood in terms of the circuit depth which increases as the time increases.

As explained in section~\ref{subsec:trotter}, the number of gates and circuit depth can be reduced by utilising the Cartan decomposition of two-qubit gates in terms of minimum number of CNOT gates. Under this scheme, the quantum simulation results labeled as ``Cartan" for a two-neutrino system are given in Figure~\ref{fig:2N_cartan}. Similar to the results with trotterization, in the case of two neutrino system, the quantum simulation results match exactly with the classical ones at initial times. With the increase in time, due to increase in circuit depth, the simulation results deviate from the classical ones. However, there are no significant differences between the ``Trotter" and ``Cartan" results, because the length of the circuits in the case of two-neutrinos is still comparable. The main advantage of ``Cartan" over ``Trotter" will be seen on the real quantum devices due to the hardware noise.

With an increase in the system size, {\it i.e.,} a four-neutrino system, the quantum simulation results for ``Trotter" and ``Cartan" are shown in Figures~\ref{fig:4N_trotter} and \ref{fig:4N_cartan}, respectively. In case of trotterization for a time-dependent Hamiltonian, the results start deviating from the exact ones at earlier times as compared to the two-neutrino case. Furthermore, the deviation at the later times is more pronounced as compared to the two-neutrino case. We interrupt the simulations at $t=270\omega_{0}^{-1}$, because the computations become very time-consuming as time increases.

In the case of four-neutrino system with the quantum algorithm with Cartan decomposition of two-qubit gates in terms of the minimum number of CNOT gates, the results shown in Figure~\ref{fig:4N_cartan} exhibit more deviations as compared to the two-qubit gates starting at the earlier times. However, similar to the two-neutrino case, even with the reduction in the number of gates in ``Cartan" algorithm, the results with both the methods {\it i.e.,} ``Trotter" and ``Cartan" look comparable. Since we are using a simulator instead of real quantum devices, the true advantage of this reduction in the number of gates will be visible when the qubits are error prone and the gates have hardware noise which occurs in NISQ devices. 
\begin{figure}[h!]
    \centering
    \includegraphics[width=0.9\linewidth]{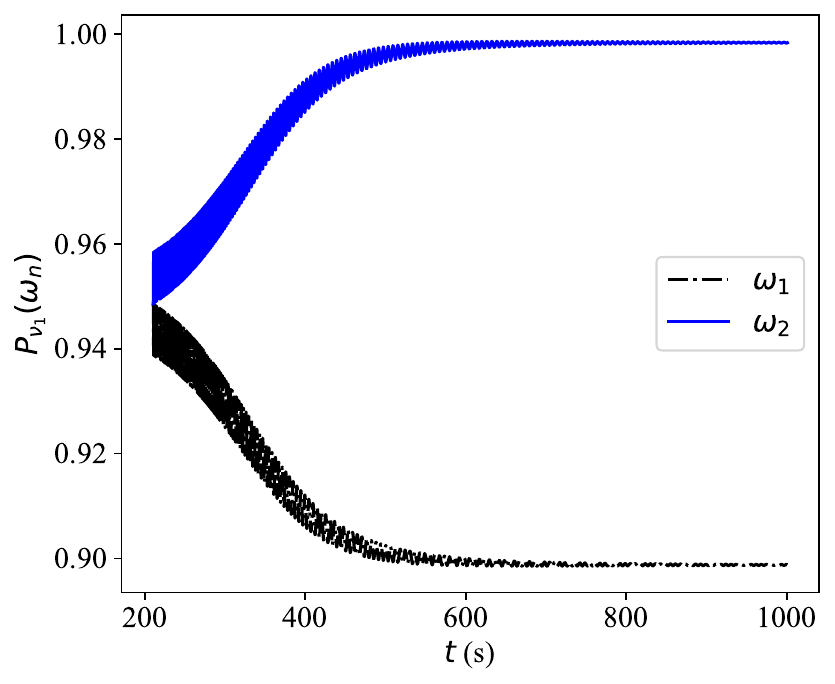}
    \caption{The survival probability for two neutrino system with an initial state $\ket{\nu_{e}^{\otimes 2}}$.}
    \label{fig:neut_2q}
\end{figure}
\begin{figure}
    \includegraphics[width=0.9\linewidth]{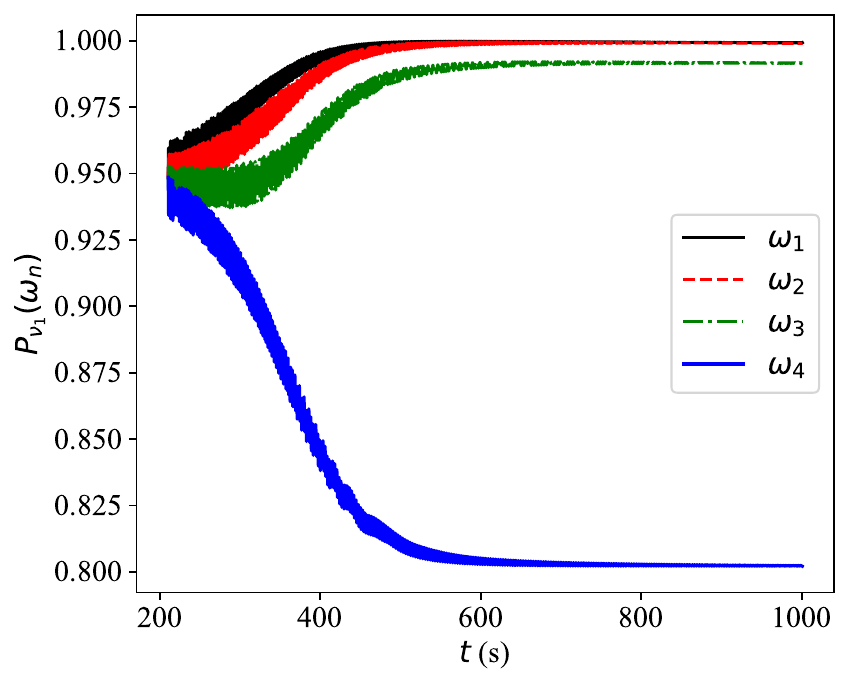}
    \caption{The survival probability for four neutrino system with an initial state $\ket{\nu_{e}^{\otimes 4}}$.}
    \label{fig:neut_2q_icon1}
\end{figure}


As can be seen from the results shown in Figures~\ref{fig:2N_trotter}, \ref{fig:2N_cartan}, \ref{fig:4N_trotter}, and \ref{fig:4N_cartan}, the quantum simulations of the collective neutrino oscillations for a large enough system are very difficult with the pure quantum algorithm. To utilize the NISQ devices for this problem to their full potential, we perform the simulations with the hybrid quantum-classical algorithm explained in section~\ref{sec:QAS}. The results corresponding to the two-neutrino and four-neutrino systems are given in Figures~\ref{fig:neut_2q} and \ref{fig:neut_2q_icon1}, respectively. The errors in the quantum simulations results are negligible and hence cannot be seen in the figures. The results match exactly with classical ones which are not shown in the plots. With this hybrid algorithm, we could perform the simulations up to very long time with accuracy.

Since only a small part of the calculations {\it i.e.,} the computations of matrix elements is performed on the quantum computer, this algorithm is very suitable for the NISQ devices. However, in future when the quantum devices become error proof, the second step of algorithm {\it i.e.,} solving Eq.~\eqref{eq:ed}, can also be performed on the quantum computer making the algorithm fully quantum.

\section{Conclusions}\label{sec:conclusions}
We have performed the quantum simulations of the collective neutrino oscillations with a time-dependent Hamiltonian using a hybrid algorithm. The generalisation of the trotterisation to the time-dependent Hamiltonian has been employed for the time evolution of the collective neutrino oscillations. It has further been optimised by performing the Cartan decomposition of two-qubit gates in the minimum number of CNOT gates. The number of gates and the circuit depth increase drastically for each time step as the time progresses in both these algorithms. However, in the later case, as the number of CNOT gates reduces approximately by half, the algorithms becomes more noise resilience and it makes it a better choice for the NISQ quantum devices. However, it still has significant deviations from the exact results.

We have employed the hybrid quantum-classical algorithm based on the quantum assisted simulator technique
to simulate the time-evolution of collective neutrino oscillations. This algorithm does not have any classical feedback loop and the number of quantum gates and circuit depth is reduced significantly. Therefore, the quantum simulations could be performed for sufficient time interval with negligible errors. 

The utilisation of the quantum computer in the hybrid quantum-classical algorithm used in the present work is minimal. Only matrix elements are calculated on the quantum computer and these are further used in simulating the time-evolution on classical computer. This limited usage makes this algorithm useful for the NISQ devices. However, further improvements can be made to increase the employment of quantum computers.
\begin{acknowledgments}

This work was supported in part by the U.S.~Department of Energy, Office of Science, Office of High Energy Physics, under Award  No.~DE-SC0019465 and in part by the National Science Foundation Grant  PHY-2108339. Kaytlin Harrison was also supported by an Open Quantum Initiative Undergraduate Fellowship at the University of Wisconsin, Madison. This work was performed under the auspices of the U.S. Department of Energy by Lawrence Livermore National Laboratory under Contract No. DE-AC52-07NA27344. This work was supported in part by the NUCLEI SciDAC-4
collaboration DE-SC001822.

\end{acknowledgments}





\newpage
\phantom{i}
\newpage
\bibliographystyle{apsrev}
\bibliography{ref}

\end{document}